# Spin Dynamics in Rotating Quantum Plasmas: Coupled EPI Dispersion and Solitary wave analysis

Atherv Saxena[1] and Punit Kumar[1*]

[1]*Department of Physics, University of Lucknow, India*

[1*]*E-mail- kumar_punit@lkouniv.ac.in*

## Abstract

The propagation of an electrostatic wave in three component e-p-i astrophysical quantum plasma in a rotating frame has been studied, taking into account the particle's spin, Fermi pressure and quantum Bohm potential. Spin polarization plays a key role in explaining the dynamics of quantum plasmas, especially in astrophysical contexts due to the high external magnetic field prevalent in such environments. Effects specific to this particular environment like rotation as well as gravity have also been included. Coupled dispersion of electron, positron and ion modes have been obtained. Further, the investigation of solitary wave by Korteweg-de-Vries method has been carried out and soliton solution has been obtained. Quantum effects increase wave dispersion and soliton stability in quantum plasma, thereby affecting the electrostatic potential.

## 1. Introduction

Quantum plasmas have gained significant attention due to their unique attributes, which arise from quantum mechanical effects. These effects become prominent when the de-Broglie wavelength of charged particles approaches inter-particle spacing [1-3]. Quantum plasmas are present in various environments, ranging from dense astrophysical settings to laboratory conditions. These include active galactic cores [4], pulsar magnetospheres [5], black hole accretion discs [6], white dwarf atmospheres [7, 8], the Van Allen radiation belts [9], the early universe [10], and the center of our galaxy [11]. In such environments, quantum effects become critical due to the high density and rotation, making quantum plasma phenomena an important subject of study [12, 13].

In astrophysical contexts, plasmas often experience rapid rotation, particularly in objects like white dwarfs, neutron stars, and pulsars [14]. As these objects collapse, their moment of inertia decreases, leading to rapid rotation due to the conservation of angular momentum. This rotation, coupled with the conservation of magnetic flux and strong magnetic fields, results in the formation of highly dense, rotating, magnetized plasma [15]. The Coriolis force in these rotating environments can create an effect analogous to a magnetic field, further complicating the plasma dynamics [16].

A subset of these plasmas is electron-positron-ion (e-p-i) plasmas, where positrons, due to their equal mass and charge magnitude to electrons, introduce unique dynamics compared to typical electron-ion systems [17-21]. The study of electrostatic waves in rotating quantum e-p-i plasma is critical, especially as ion-acoustic waves (IAWs) and magnetosonic waves become central to understanding wave propagation in these systems [22-25]. Ion temperature, pressure degeneracy, and exchange-correlation effects have been explored in the context of IAWs in e-p-i plasma, revealing interesting behaviors [26-30].

The Quantum Hydrodynamic (QHD) model has been employed to describe these systems, offering a computationally efficient approach by using macroscopic variables and simplifying the handling of boundary conditions [31,32]. However, early hydrodynamic models for quantum particles, which treated the evolution of spin-up and spin-down states alike, were found to violate the Pauli Exclusion Principle [33]. The standard QHD model, while effective, does not account for the spin polarization of particles, which limits its accuracy in predicting the behaviour of quantum plasmas. This limitation is addressed by, the Separated Spin Evolution Quantum Hydrodynamic (SSE-QHD) model which addresses spin-up and spin-down particles as separate species of particles [34]. New wave modes have been obtained using this model [35-37]. Spin effects also help in better understanding the magnetic field associated with plasma waves, which is crucial in both laboratory experiments and astrophysical applications [38,39]. The SSE-QHD model has thus become a pivotal tool in the study of quantum plasma, offering a more comprehensive framework than earlier models [40-56].

This paper is devoted to the analysis of coupled dispersion relation for multi-component plasmas and the investigation of solitary structures. Section – 2, describes the basic set of equations of the SSE-QHD model and the particle dynamics. Section - 3 is devoted to the study of coupled dispersion of electron, positron and ion in a rotating magnetized e – p - i quantum plasma using the QHD model. Section - 4, is devoted to the investigation of solitary waves by applying the KdV method and it's soliton solutions. Finally, section - 5 presents summary and discussion.

## 2. Quantum Plasma dynamics

We consider a collisionless electron-positron-ion quantum plasma, in the presence of constant external magnetic field along the z direction $\vec{B} = B_0(\hat{z})$, and an external electrostatic wave $\vec{E} = -\nabla\phi$, where $\phi$ is the electrostatic potential interacts with the plasma. The plasma is assumed to be rotating in the astrophysical settings with angular frequency $\vec{\Omega}$, at an angle $\theta$ to the direction of the magnetic field.

We perform linear analysis by considering the electrostatic wave to propagate obliquely to the external magnetic field in the x-z plane, i.e., $\vec{\nabla} = \left(\frac{\partial}{\partial x}, 0, \frac{\partial}{\partial z}\right)$, $k = \left(k_\perp, 0, k_\parallel\right)$ and $\vec{\Omega} = (\Omega_x, 0, \Omega_z)$, with $\Omega_x = \Omega_0 \sin\theta$ and $\Omega_z = \Omega_0 \cos\theta$. All the perturbations take the form $\sim \exp(ik_\perp + ik_\parallel - i\omega t)$ where $k\left(=\sqrt{k_\perp^2 + k_\parallel^2}\right)$ is the propagation vector.

### 2.1. Fermion dynamics

The momentum and continuity equations for fermions, considering spin-up and spin-down states as separated species of particles, with the subscript $j = e, p$ denoting the electron and positron, respectively, are

$$0 = \frac{q_j}{m_j}\left[\vec{E} + \left(\vec{v}_{j\alpha} \times \vec{B}\right)\right] - \left(\frac{1}{m_j n_{j\alpha}}\vec{\nabla}P_{j\alpha}\right) + \frac{\hbar^2}{2m_j^2}\vec{\nabla}\left(\frac{1}{\sqrt{n_{j\alpha}}}\vec{\nabla}^2\sqrt{n_{j\alpha}}\right) + 2m_j\left(\vec{v}_{j\alpha} \times \vec{\Omega}\right), \quad (1)$$

$$\frac{\partial n_{j\alpha}}{\partial t} + \vec{\nabla}\cdot\left(n_{j\alpha}\vec{v}_{j\alpha}\right) = 0, \quad (2)$$

where, $v_j$, $m_j$, $n_j$, $P_j$ represent the fluid velocity, rest mass, particle density and Fermi pressure of the $j^{th}$ species of particles and $\alpha = \uparrow$ and $\downarrow$ denotes spin-up and spin-down fermions respectively. The left hand side of equation (1) is taken 0 because the fermions are considered inetia-less with respect to ions, as they are much lighter than ions [35]. The first term, on the right hand side of equation (1) refers to the Lorentz force, the second term is the force due to the degenerate pressure [44] $P_{j\alpha}(=m_j V_{Fj\alpha}^2 n_{j\alpha}^{5/3}/5n_{oj}^{2/3})$, where $V_{Fj\alpha} = \sqrt{\zeta_\alpha} V_{Fj}$ is the Fermi velocity of fermions with $V_{Fj}(=\hbar(3\pi^2 n_{0j})^{1/3}/m_j)$ and $\zeta_\alpha = \left[(1-\eta)^{5/3} + (1+\eta)^{5/3}\right]/2$, and $\eta(=\Delta n_{j\alpha}/n_0)$ is the spin polarization due to the presence of magnetic field, $\Delta n_{j\alpha} = \sum_j (n_\uparrow - n_\downarrow)$ denotes the concentration difference of spin-up $(\uparrow)$ and spin-down $(\downarrow)$ fermions. The third term is the quantum Bohm force involving quantum electron tunneling in dense quantum plasma [45]. The last term, is the Coriolis force, due to the rotation of the plasma with angular velocity $\vec{\Omega}$. Since, rotation is taken to be slow, quadratic and higher order terms such as centrifugal force $\vec{\Omega} \times (\vec{\Omega} \times \vec{r})$ are safely neglected [15].

Perturbatively expanding eqs. (1) and (2) in orders of the fields associated with the external electrostatic wave and assuming all the varying parameters to take the form,

$$f = f_0 + f^{(1)}$$

with $f_0$ representing the initial value, and $f^{(1)}$ is the perturbation term. The first order momentum and continuity eqs. for fermions now become

$$0 = \frac{q_j}{m_j}\left[\vec{E} + \left(\vec{v}_{j\alpha}^{(1)} \times \vec{B}_0 - \vec{v}_{0j} \times \vec{B}^{(1)}\right)\right] - \left(\frac{1}{m_j n_{j\alpha}}\vec{\nabla} P_{j\alpha}\right) + \frac{\hbar^2}{2m_j^2}\vec{\nabla}\left(\frac{1}{\sqrt{n_{j\alpha}}}\vec{\nabla}^2\sqrt{n_{j\alpha}}\right) + 2m_j\left(\vec{v}_{j\alpha}^{(1)} \times \vec{\Omega}_0 - \vec{v}_{0j} \times \vec{\Omega}^{(1)}\right),$$

(3)

and

$$\frac{\partial n_{j\alpha}^{(1)}}{\partial t} + n_{0j}\vec{\nabla}\cdot\left(\vec{v}_{j\alpha}^{(1)}\right) = 0.$$

(4)

After performing the necessary algebra, we arrive at the spatial components of perturbed velocity of the electron,

$$\vec{v}_{ex\alpha}^{(1)} = \left[\Omega_{eff} k_{\Box} + 2\Omega_x k_{\perp}\right]\frac{2\omega\Omega_x}{\Gamma_{Qe\alpha}^{\Box}\left(k_{\Box}\Omega_{eff}\right)^2 + 4\Gamma_{Qe\alpha}^{\perp}\left(k_{\perp}\Omega_x\right)^2}\phi^{(1)},$$

(5)

$$\vec{v}_{ez\alpha}^{(1)} = \left[\Omega_{eff} k_{\Box} + 2k_{\perp}\Omega_x\right]\frac{\Omega_{eff}\omega}{\Gamma_{Qe\alpha}^{\Box}\left(k_{\Box}\Omega_{eff}\right)^2 + 4\Gamma_{Qe\alpha}^{\perp}\left(k_{\perp}\Omega_x\right)^2}\phi^{(1)}.$$

(6)

Similarly, the spatial components of the perturbed velocity of positron are,

$$\vec{v}_{px\alpha}^{(1)} = \left[\Omega_{eff} k_{\Box} - 2\Omega_x k_{\perp}\right]\frac{2\omega\Omega_x}{\Gamma_{Qp\uparrow\downarrow}^{\Box}\left(k_{\Box}\Omega_{eff}\right)^2 - 4\Gamma_{Qp\uparrow\downarrow}^{\perp}\left(k_{\perp}\Omega_x\right)^2}\phi^{(1)},$$

(7)

$$\vec{v}_{pz\alpha}^{(1)} = \left[\Omega_{eff} k_{\Box} - 2k_{\perp}\Omega_x\right]\frac{\Omega_{eff}\omega}{\Gamma_{Qp\uparrow\downarrow}^{\Box}\left(k_{\Box}\Omega_{eff}\right)^2 - 4\Gamma_{Qp\uparrow\downarrow}^{\perp}\left(k_{\perp}\Omega_x\right)^2}\phi^{(1)}.$$

(8)

where, $\Omega_{eff} = \omega_{cj} + 2\Omega_0\cos\theta$ is the effective angular frequency due to rotation, in terms of cyclotron frequency of fermions $\omega_{cj} = eB_0/m_j$. The term $\left[\Omega_{eff} k_{\Box} \pm 2\Omega_x k_{\perp}\right]$ represents the combined effects of wave propagation and plasma rotation on fermion velocity, while the term

$\Gamma_{Qp\uparrow\downarrow}^{\Box}\left(k_{\Box}\Omega_{eff}\right)^2 \pm 4\Gamma_{Qp\uparrow\downarrow}^{\perp}\left(k_{\perp}\Omega_x\right)^2$ reflects the quantum coupling that modulates the fermion's velocity influenced by quantum corrections like Fermi pressure, Bohm potential and spin polarization.

### 2.1.2. Ion dynamics

In the case of ions, due to their large mass as compared to the electrons and positrons, i.e., $m_{e,p}/m_i \ll 1$, the quantum effects are insignificant and so they can be considered classical [35]. The governing equations for ions in the rotating frame of reference are,

$$\frac{\partial n_i}{\partial t} + \vec{\nabla}\cdot\left(n_i \vec{v}_i\right) = 0, \tag{9}$$

$$m_i\left(\frac{\partial \vec{v}_i}{\partial t} + (\vec{v}_i \cdot \nabla)\vec{v}_i\right) = e\left[\vec{E} + \left(\vec{v}_i \times \vec{B}\right)\right] - m_i \nabla \Phi + 2m_i\left(\vec{v}_i \times \vec{\Omega}\right). \tag{10}$$

In the above equations $v_i$, $m_i$, $n_i$ represent the fluid velocity, rest mass and particle density of ion. The second term in the momentum equation, is the gravitational potential term which is derived from the Poisson's equation for gravitational potential field as, $\nabla^2 \Phi_j = 4\pi G n_j$ [54], where $G$ is the gravitational constant.

Perturbatively expanding eqs. (3) and (4) in orders of the fields as done in previous section, the momentum and continuity equations for ion are,

$$\frac{\partial \vec{v}_i^{(1)}}{\partial t} = \frac{e}{m_i}\vec{E}^{(1)} + \frac{e}{m_i}\left(\vec{v}_i^{(1)} \times \vec{B}_0 - \vec{v}_{0i} \times \vec{B}^{(1)}\right) - \nabla \varphi^{(1)} + 2\left(\vec{v}_i^{(1)} \times \vec{\Omega}_0 + \vec{v}_{0i} \times \vec{\Omega}^{(1)}\right), \tag{11}$$

and

$$\frac{\partial n_i^{(1)}}{\partial t} + n_{0i} \nabla \cdot \left( \vec{v}_i^{(1)} \right) = 0. \tag{12}$$

The spatial components of ion velocity are found to be,

$$\vec{v}_{ix}^{(1)} = \left[ \left( \Gamma_{G\Box} - 4\Omega_x^2 \right) k_\perp + 2\Omega_{eff} \Omega_x k_\Box \right] \frac{\omega}{\left( \Gamma_{G\Box} - 4\Omega_x^2 \right) \Gamma_{G\perp} - \Gamma_{G\Box} \Omega_{eff}^2} \phi^{(1)}, \tag{13}$$

$$\vec{v}_{iz}^{(1)} = \left[ k_\Box \left( 1 - \frac{2i\Omega_x}{\Gamma_{G\Box} - 4\Omega_x^2} \right) + \frac{2\Omega_x \Gamma_{G\Box} \Omega_{eff}}{\left( \Gamma_{G\Box} - 4\Omega_x^2 \right) \Gamma_{G\perp} - \Gamma_{G\Box} \Omega_{eff}^2} \left( k_\perp + \frac{2\Omega_{eff} \Omega_x k_\Box}{\Gamma_{G\Box} - 4\Omega_x^2} \right) \right] \frac{\omega}{\Gamma_{G\Box}} \phi^{(1)}. \tag{14}$$

where, $\Gamma_{G\Box,\perp} = \omega^2 - n_{0i} \rho k_{\Box,\perp}^2$ is the parameter arising due to gravitational effects. In both of the above equations, term in the denominator $\left( \Gamma_{G\Box} - 4\Omega_x^2 \right)$ occur due to the critical balance of the plasma's rotation frequency with the gravitational potential, causing resonances that can alter the ion motion. This represent conditions where the system's stability is highly sensitive, and small perturbations can result in large, potentially unstable responses in the quantum plasma.

## 3. Dispersion Relation of coupled EPI mode

For homogenously magnetized quantum plasma in the presence of electrons, positrons and ions, the quasi-neutrality condition gives, $n_e + n_p \cong n_i$ and in equilibrium we have $n_{0e} + n_{0p} = n_{0i}$.

The system is closed by Poisson's equation,

$$\nabla^2 \phi = \zeta_\alpha n_{e\alpha} + \zeta_\alpha n_{p\alpha} - n_i. \tag{15}$$

By applying the linear analysis and neglecting convective time derivatives, we arrive at the following dispersion relation

$$k^2 = \frac{\zeta_\alpha n_{0e}\left(\Omega_{eff}^{(e)} k_\Box + 2\Omega_x k_\perp\right)^2}{\Gamma_{Qe\alpha}^\Box \left(\Omega_{eff}^{(e)} k_\Box\right)^2 + \Gamma_{Qe\alpha}^\perp \left(2k_\perp \Omega_x\right)} - \frac{\zeta_\alpha n_{0p}\left(\Omega_{eff}^{(p)} k_\Box - 2\Omega_x k_\perp\right)^2}{\Gamma_{Qp\alpha}^\Box \left(\Omega_{eff}^{(p)} k_\Box\right)^2 + \Gamma_{Qp\alpha}^\perp \left(2k_\perp \Omega_x\right)} - n_{0i}\left[\frac{k_\perp^2}{\Gamma_{G\perp}} + \frac{k_\Box^2}{\Gamma_{G\Box}}\left(\frac{\Gamma_{G\Box} - 2i\Omega_x}{\Gamma_{G\Box} - 4\Omega_x^2}\right)\right]$$

(16)

where, $\Gamma_{Qe\alpha}^{\Box,\perp} = V_{Fe\alpha}^2 \left[\frac{(2\zeta_\alpha)^{2/3}}{5} + \frac{H_e^2 k_{\Box,\perp}^2}{4\omega_{pe}^2}\right]$ is the quantum coupling parameter for electrons with

$H_{j\alpha}\left(=\frac{\hbar \omega_{pj}}{m_j V_{Fe\alpha}}\right)$ being the dimensional quantum parameter which is a ratio of the plasmon energy (the energy of an elementary excitation associated with fermion plasma wave) to the kinetic energy, representing the quantum diffraction effects [44] and $\omega_{pj}\left(=\frac{n_{0j}e^2}{m_j \varepsilon_0}\right)^{1/2}$ is the plasma frequency for fermions, where $\varepsilon_0$ is the electric permittivity of free space. In the limit $G \to 0$, the above dispersion relation will fold into the dispersion relation for non gravitating quantum plasmas.

The first term in the R.H.S of the above dispersion relation is responsible for the contribution of higher frequency electron wave and the second term is responsible for the contribution of positron mode. The third term comes from the contribution of ion mode. Numerical investigations have adopted astrophysical plasma conditions typical of the outer layers of compact stars like white dwarfs or neutron stars. These parameters include ion number density $10^{26} m^{-3} \leq n_{0i} \leq 10^{28} m^{-3}$, where pair annihilation effects are negligible in such dense electron-positron plasmas, magnetic field strength $B_0 = 10 - 10^5 T$, and $T_{Fe} \approx 10^7 - 10^8 K$ so that $0.24 \leq H \leq 1.10$ [35-40].

Figure 1 shows the variation normalised propagation vector $kc/\omega_p$, which represents energy or power transmission direction and magnitude with $\omega/\omega_p$ (transmissibility of wave in plasma). As the wave's angular frequency increases, the magnitude of the propagation vector decreases, indicating an inverse proportionality between dispersion and frequency. This implies that lower angular frequencies correspond to higher photon energies, which leads to increased plasma interaction and faster energy loss as wave energy rises. This trend is evident in both quantum plasma as well as in the absence of quantum effects $(\hbar \to 0)$, though the latter shows a 53% lower magnitude of this effect. This discrepancy arises due to the substantial Fermi pressure in quantum plasma compared to the thermal pressure, leading to an increased number of accessible energy levels and a higher state density in the plasma. Consequently, the plasma's ability to transmit waves is constrained, resulting in a noticeable reduction in angular frequency as the propagation vector grows.

Figure 2 shows the variation of $kc/\omega_p$ with $\omega/\omega_p$ for different values of quantum parameter $H$. Across all three cases, a similar trend is observed. As the quantum parameter decreases, the value of the propagation vector increases by 33.3% in the first case and 23% in the next case. This phenomenon arises due to the interaction between degeneracy pressure and the quantum Bohm potential. As a result, we can conclude that wave transmission becomes constrained by the influence of the quantum Bohm potential, which incorporates quantum tunneling effects.

Figure 3 displays the variation in normalised propagation vector $kc/\omega_p$ with wave frequency $\omega/\omega_p$ for varying degrees of spin-polarization. The solid, dashed, and dotted lines correspond to the propagation vector variations for fully spin-polarized (η = 0.81), partially polarized (η = 0.008), and unpolarized (η = 0) plasmas, respectively. Notably, wave power transmission increases with greater spin-polarization and stabilizes at higher values of the propagation vector. Specifically, for the fully spin-polarized case, the propagation vector exceeds that of the unpolarized plasma by approximately 83% around k ≈ 0.8. This discrepancy is due to the increased Fermi pressure resulting from spin-polarization and the electron's spin magnetic moment, both of which are crucial in the presence of a magnetic field.

## 4. Analysis of Ion Acoustic Solitary wave

In this section, we examine the nature of an obliquely propagating ion acoustic wave in a magnetized quantum plasma, the standard perturbation technique [35-37] has been adopted.

The space and time variables are stretched as,

$$\xi = \varepsilon^{1/2}\left(l_x x + l_z z - Vt\right), \quad \tau = \varepsilon^{3/2} t \tag{17}$$

where, $\varepsilon$ is a parameter which determines the strength of nonlinearity, $l_x$ and $l_z$ are the direction cosines of the wave vector $k$ along the x and z axes, respectively such that $l_x^2 + l_z^2 = 1$ and $V$ is the phase velocity of the wave. Now, expanding the perturbed quantities $n_i$, $v_{ix}$, $v_{iy}$, $v_{iz}$, and $\phi$ in terms of $\varepsilon$, we get,

$$n_i = 1 + \varepsilon n_i^{(1)} + \varepsilon^2 n_i^{(2)} + ...$$

$$v_{ix} = \varepsilon^2 v_{ix}^{(1)} + \varepsilon^3 v_{ix}^{(2)} + ...$$

$$v_{iy} = \varepsilon^{3/2} v_{iy}^{(1)} + \varepsilon^{5/2} v_{iy}^{(2)} + ...$$

$$v_{iz} = \varepsilon v_{iz}^{(1)} + \varepsilon^2 v_{iz}^{(2)} + ...$$

$$\phi = \varepsilon \phi^{(1)} + \varepsilon^2 \phi^{(2)} + ... \tag{18}$$

The strong magnetic field in the z direction and the plasma rotation introduce an anisotropy in the velocity components $v_{ix,iy,iz}$. Substituting the above perturbation scheme in the equations (1) to (5), we obtain,

$$v_{iz}^{(1)} = \frac{V}{l_z} n_i^{(1)}, \tag{19}$$

$$v_{iy}^{(1)} = \frac{l_x}{\Omega_{eff}^{(i)}} \frac{\partial \phi^{(1)}}{\partial \xi} + V \frac{\partial \Phi}{\partial \xi}, \tag{20}$$

$$v_{ix}^{(1)} = \frac{V}{\Omega_{eff}^{(i)}} \frac{\partial v_{iy}^{(1)}}{\partial \xi} + 4\pi G \frac{\partial \Phi}{\partial \xi}, \tag{21}$$

$$V \frac{\partial v_{iz}^{(1)}}{\partial \xi} = l_z \frac{\partial \phi^{(1)}}{\partial \xi} + \alpha \sin \theta v_{iy}^{(1)}. \tag{22}$$

The, lower order terms of Poisson equation gives a relation between densities of plasma species as,

$$\zeta_\alpha n_{e\alpha}^{(1)} + \zeta_\alpha n_{p\alpha}^{(1)} = n_i^{(1)} \tag{23}$$

The coefficient of $\varepsilon^{3/2}$ order terms of spatial components of momentum equation for fermions are,

$$\frac{\partial \phi^{(1)}}{\partial \xi} = \frac{(2\zeta_\alpha)^{2/3}}{5} \frac{\partial n_{e\alpha}^{(1)}}{\partial \xi} - 2\left(\Omega_x \frac{\partial v_{jx\alpha}^{(1)}}{\partial \tau}\right), \tag{24}$$

Using eqs. (13-18), we get the phase velocity of ion acoustic wave,

$$V = \frac{2^{1/3}}{5\left(\zeta_\uparrow^{1/3} + \zeta_\downarrow^{1/3}\right)} \left(l_z^2 + \frac{l_z l_x \alpha \sin\theta}{\Omega_{eff}}\right). \tag{25}$$

Collecting higher order terms from the continuity equation for ions, corresponding to $\varepsilon^{5/2}$ orders

$$V \frac{\partial n_i^{(2)}}{\partial \xi} - l_z \frac{\partial v_{iz}^{(2)}}{\partial \xi} = \frac{\partial n_i^{(2)}}{\partial \tau} + l_x \frac{\partial v_{ix}^{(1)}}{\partial \xi} + l_z \frac{\partial \left(n_i^{(1)} v_{iz}^{(1)}\right)}{\partial \xi}. \tag{26}$$

From the spatial components of momentum equations (13 and 14) for ions, the terms corresponding to $\varepsilon^{5/2}$ orders are,

$$l_x \frac{\partial \phi^{(2)}}{\partial \xi} = V \frac{\partial v_{ix}^{(2)}}{\partial \xi} + \alpha v_{iy}^{(2)}, \tag{27}$$

$$V \frac{\partial v_{iy}^{(2)}}{\partial \xi} = -\Omega_{eff}^{(i)} v_{ix}^{(2)} + 4\pi G \frac{\partial n_{iz}^{(2)}}{\partial \xi}, \tag{28}$$

$$V \frac{\partial v_{iz}^{(2)}}{\partial \xi} - l_z \frac{\partial \phi^{(2)}}{\partial \xi} + \cos\theta v_{iy}^{(2)} = \frac{\partial v_{iz}^{(1)}}{\partial \tau} + l_z v_{iz}^{(1)} \frac{\partial v_{iz}^{(1)}}{\partial \xi}. \tag{29}$$

The higher order term of Poisson's equation corresponding to $\varepsilon^2$ orders is,

$$\frac{\partial^2 \phi^{(2)}}{\partial \xi^2} = \zeta_\alpha n_{e\alpha}^{(2)} + \zeta_\alpha n_{p\alpha}^{(2)} - n_i^{(2)}. \tag{30}$$

The next higher order terms of the spatial components of momentum equation for fermions having spin-up and spin-down state are,

$$\frac{\partial \phi^{(2)}}{\partial \xi} - \frac{(2\zeta_\alpha)^{2/3}}{5} \frac{\partial n_{j\alpha}^{(2)}}{\partial \xi} = \frac{(2\zeta_\alpha)^{2/3}}{15} n_{j\alpha}^{(1)} \frac{\partial n_{j\alpha}^{(1)}}{\partial \xi} - l_z^2 \Gamma_{Qj\alpha}^{\Box,\bot} \frac{\partial n_{j\alpha}^{(1)}}{\partial \xi}. \tag{31}$$

Finally, eliminating $n_i^{(2)}, \phi^{(2)}, v_{iy}^{(2)}, v_{iz}^{(2)}$ from equations (26-31), we obtain the KdV equation,

$$\frac{\partial \phi^{(1)}}{\partial \tau} + C_1 \phi^{(1)} \frac{\partial \phi^{(1)}}{\partial \xi} + C_2 \frac{\partial^3 \phi^{(1)}}{\partial \xi^3} = 0, \tag{32}$$

where, the nonlinear and dispersive coefficients $C_1$ and $C_2$ respectively, are

$$C_1 = \frac{5}{2} l_z (1 - \zeta_\alpha), \tag{33}$$

$$C_2 = \frac{V^2}{2l_z} \left( \left( \frac{l_x^2 + l_x l_z \alpha \sin\theta}{\Omega_{eff}^2} \right) \left( \frac{V\Omega_{eff}}{l_z \Omega_{eff} + l_x \alpha \sin\theta} \right) - \Gamma_{Q\alpha}^{\Box} \frac{l_z^2}{15V} (\zeta_\alpha^{1/3}) \right). \tag{34}$$

For the solution of equation (26), we use the transformation $\psi = \xi - u\tau$ for a co-moving frame with velocity $u$ of the nonlinear structure. We get the soliton solution,

$$\phi^{(1)} = \sigma \operatorname{sech}^2 \left( \frac{\psi}{\gamma} \right), \tag{35}$$

where, $\sigma = \dfrac{5u}{C_1}$ and $\gamma = \sqrt{\dfrac{4C_2}{u}}$ represents the peak amplitude and width of soliton, respectively. It is evident from the equations (33) and (34) that the amplitude of the soliton is governed by $C_1$, as this coefficient reflects how strongly the nonlinearity affects the soliton's peak, and the dispersive coefficient $C_2$ affects the width by controlling quantum dispersion and rotational effects contribute to the soliton's spreading behavior.

Figure 4 shows a comparison of soliton profiles in quantum and corresponding classical case $(\hbar \to 0)$ with the variation of stretched coordinate ($\xi$) with electrostatic potential ($\phi$). The results demonstrate a significant rise in the electrostatic potential of the soliton's profile when quantum correction terms, linked to Fermi pressure, the quantum Bohm potential, and spin effects of fermions are integrated into the quantum plasma. These quantum corrections induce energy redistribution and increased electron occupancy in higher energy states, ultimately enhancing the stability and dynamics of solitons. Consequently, there is an overall enhancement in the electrostatic potential associated with the soliton profile of 60% within the quantum plasma.

Figure 5 shows the variation of stretched coordinate ($\xi$) with electrostatic potential ($\phi$) for soliton profile for varying ion density. The electrostatic potential of the soliton profile decreases by approximately 50% with increase in ion density. As the density increases, the balance of forces within the soliton structure is altered. The increased Fermi pressure and quantum effects can lead to a redistribution of charge and energy within the soliton, causing the electrostatic potential to decrease. The Fermi pressure tends to resist compression, as the particle density increases, the Fermi pressure also increases, which resists further compression of the electron and positron clouds acting as a repulsive force that counteracts the potential.

Figure 6 is a 3D representation of soliton profile in quantum plasma with the variation of electrostatic potential ($\phi$) and the stretched coordinates $\xi$ and $\tau$. A contour of the 3D structure is also mentioned describing the stability of soliton's peak for approximately 2 seconds and soliton structure for approximately 10 seconds, in the presence of coriolis force and gravitational effects. The red colour shows the stability of peak of soliton and the blue, green and yellow colours are representing the width of soliton and the orange colour shows starting and ending phase of soliton structure.

## 5. Summary and discussion

The dynamics of uniform astrophysical magnetised quantum plasma consisting of electrons, ions, and positrons have been studied with the investigation of ion-acoustic solitary waves and finally the soliton solution has been obtained. The Separated Spin Evolution Quantum Hydrodynamic (SSE-QHD) model is introduced as the analytical framework, incorporating quantum diffraction, quantum statistical effects and separated spin-up and spin-down effects for fermions. The dispersion relation for coupled ion, electron, and positron modes considering various quantum effects and environmental influences has been established. The solitary wave structures due to ion acoustic wave and its solution by the use of KdV method have been investigated. Through theoretical formulation and graphical analysis, the paper aims to provide a better understanding of quantum plasma dynamics.

Plasma wave transmission exhibits an inverse correlation with the magnitude of the normalized propagation vector, indicating decreased transmission efficiency as vector magnitude increases due to intensified plasma interaction with higher photon energy, leading to accelerated energy loss. Quantum plasma, characterized by greater wave dispersion by 53% compared to the absence of quantum effects, attributes this phenomenon to the presence of Fermi pressure. Additionally, spin-polarization amplifies power transmission of waves, notably increasing the propagation vector by 83% in fully polarized plasma scenarios. Integration of quantum correction terms, such as Fermi pressure, quantum Bohm potential, and fermion spin effects, elevates electrostatic potential within soliton profiles in quantum plasma compared to non quantum plasma counterparts. These corrections induce energy redistribution and heightened electron occupancy in higher energy states, fortifying soliton stability and dynamics by 60%. Further, variations in ion particle density influence the electrostatic potential of soliton profiles, with increasing density correlating with decreased electrostatic potential by approx. 50%, driven by charge redistribution influenced by Fermi pressure and quantum effects.

The present study will be helpful in analysing the behaviour of electron-positron plasmas in extreme environments such as quasars, gamma-ray bursts, active galactic cores, pulsar magnetospheres, black hole accretion discs, white dwarf atmospheres, Van Allen radiation belts, also exist in the early universe, as well as the centre of our galaxy. This comprehensive framework not only offers insight into fundamental physical processes, but also facilitates the interpretation of observational data from astrophysical sources, thus advancing our understanding of the universe's most energetic and enigmatic phenomena.

## Acknowledgement

Financial support from SERB – DST under MATRICS is gratefully acknowledged (grant no. : MTR/2021/000471).

**Figure Captions**

**Figure 1:** Variation of $\omega/\omega_p$ with $kc/\omega_p$ for coupled e-p-i modes with H=1.1 (solid line) with $\theta = 4°$, $\Omega_0 = 0.0003$, $B_0 = 2\times10^4 T$ and H =0 (dashed line) with thermal velocity $V_{Te} = 10^7\ cm/\sec$.

**Figure 2:** Variation of $\omega/\omega_p$ with $kc/\omega_p$ for coupled e-p-i modes with $\theta = 4°$, $\Omega_0 = 0.0003$, $B_0 = 2\times10^4 T$, H = 1.10 (dotted line), H = 0.4 (dashed line), H = 0.2 (solid line).

**Figure 3:** Variation of $\omega/\omega_p$ with $kc/\omega_p$ for coupled e-p-i modes with $\theta = 4°$, $\Omega_0 = 0.0003$, H=1.1, $B_0 = 2\times10 T$ with $\eta = 0.81$ (dotted line), $B_0 = 2\times10^2 T$ with $\eta = 0.008$ (dashed line) and $B_0 = 2\times10^4 T$ with $\eta = 0.0008$ (solid line).

**Figure 4:** Variation of $\phi$ with $\tau$ for soliton solution l =0.2, k =0.8, H=1.1 (solid line), $\theta = 4°$, $\Omega_0 = 0.0003$, $B_0 = 2\times10^4 T$ and H =0 (dashed line)

**Figure 5:** Variation of $\phi$ with $\tau$ for soliton solution l=0.2, k =0.8, H=1.1, $\theta = 4°$, $\Omega_0 = 0.0003$, $B_0 = 2\times10^4 T$ with ion particle density $n_i = 10^{28} m^{-3}$ (solid line), $n_i = 0.5\times10^{28} m^{-3}$ (dashed line) and $n_i = 0.25\times10^{28} m^{-3}$ (dotted line).

**Figure 6:** Variation of $\phi$ with $\xi$ for soliton solution with $\tau = 0.5$, l=0.2, k =0.8, H=1.1, $\theta = 4°$, $\Omega_0 = 0.0003$, $B_0 = 2\times10^4 T$ in a 3D representation with its contour.

**Figures**

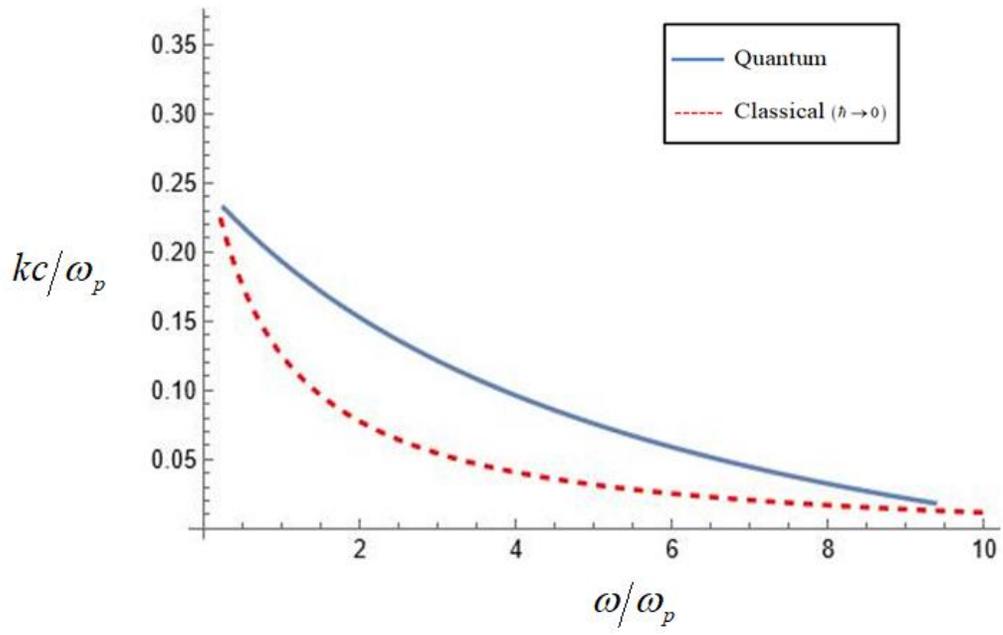

**Figure 1**

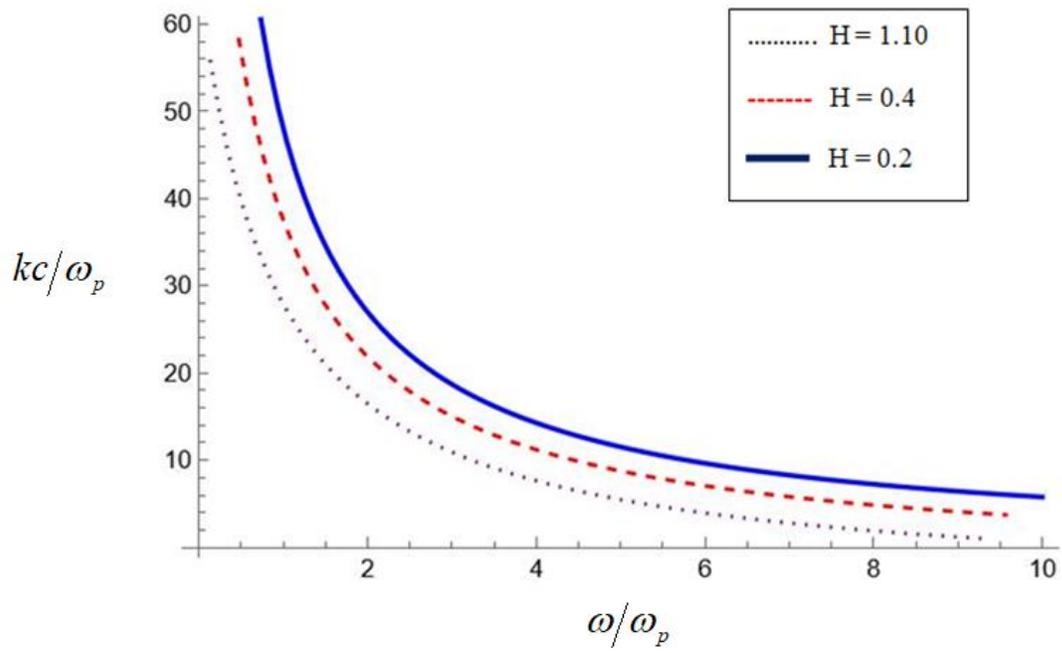

**Figure 2**

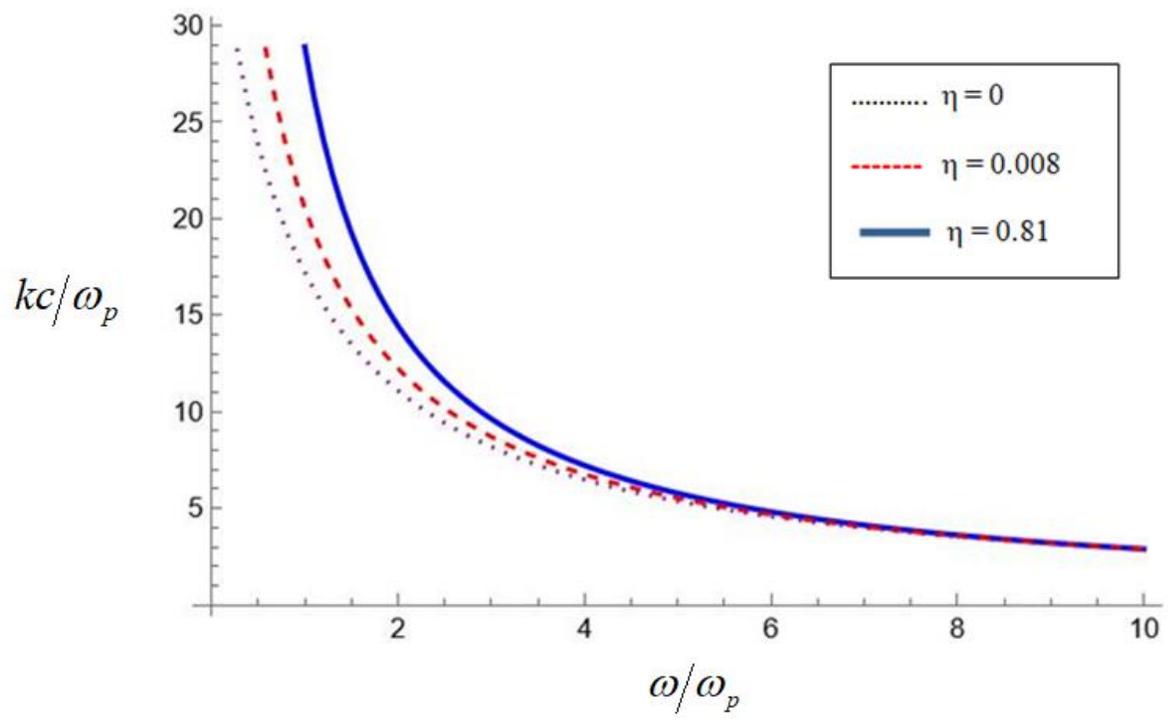

**Figure 3**

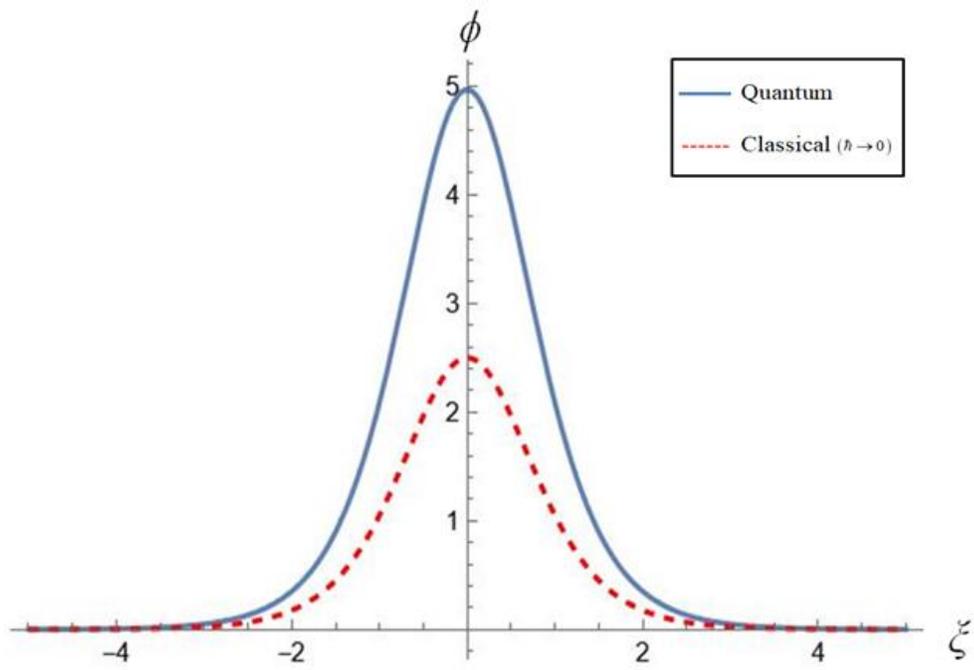

**Figure 4**

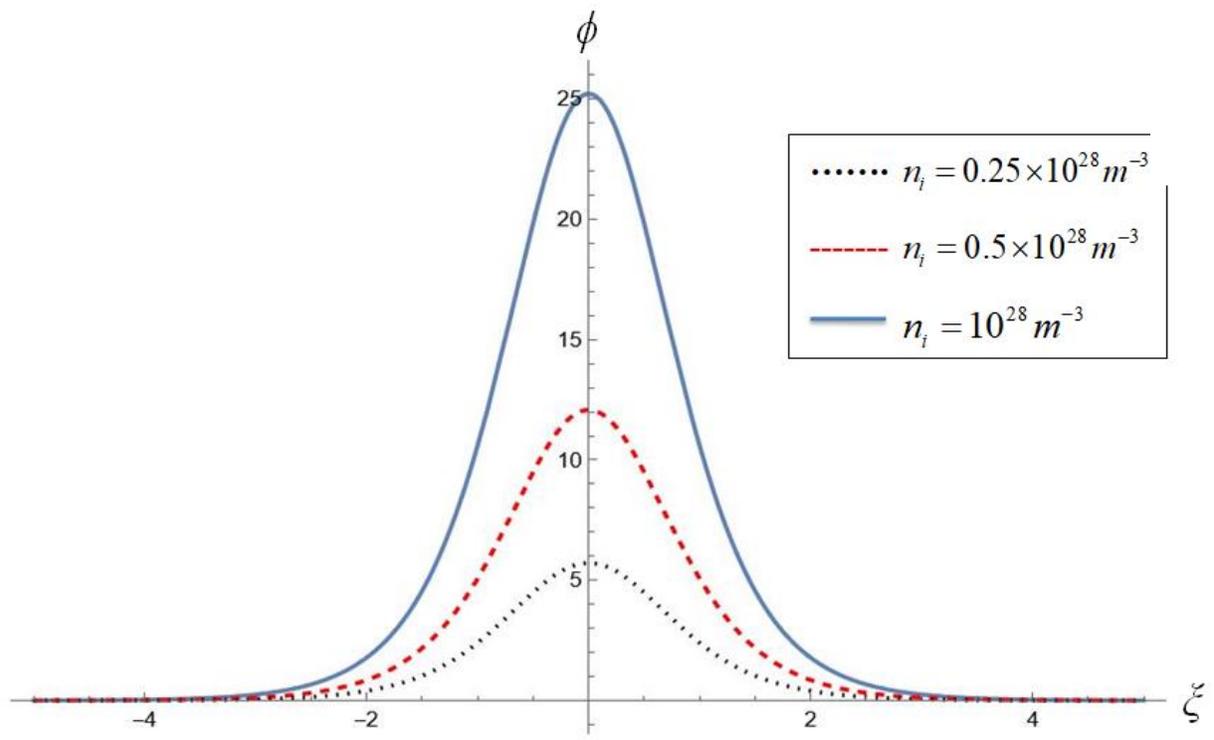

**Figure 5**

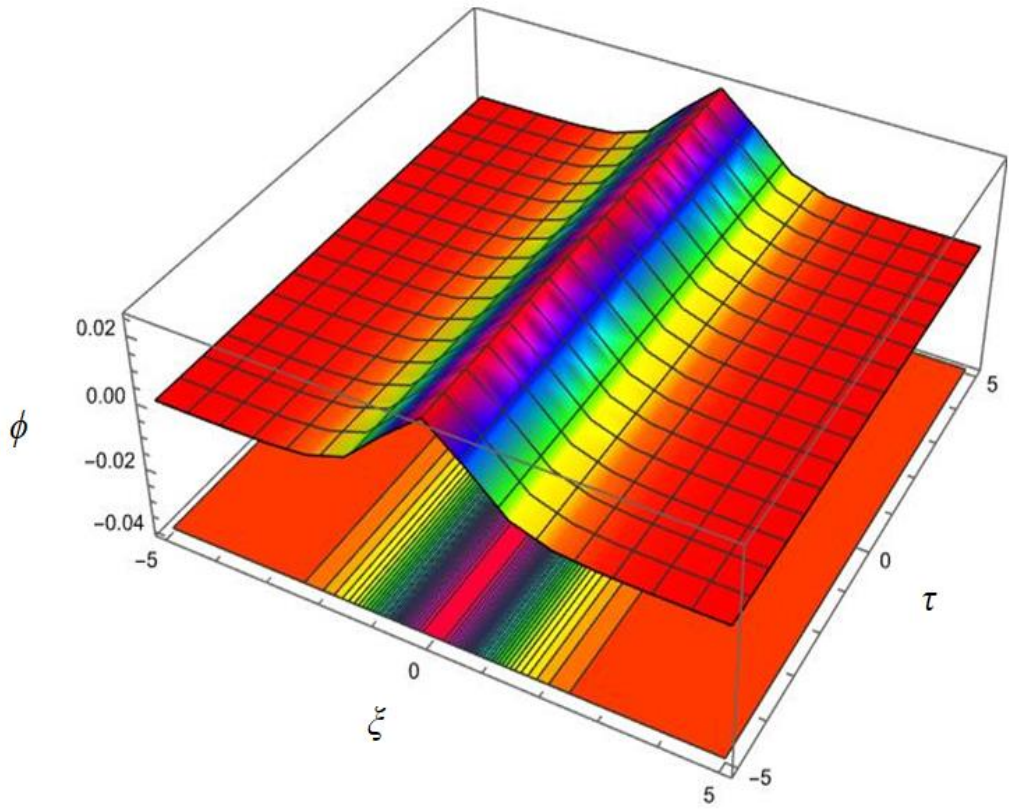

**Figure 6**